# Ultrafast symmetry control in photoexcited quantum dots

Burak Guzelturk[1,*], Joshua Portner[2], Justin Ondry[2], Samira Ghanbarzadeh[3], Mia Tarantola[4], Ahhyun Jeong[2], Thomas Field[3], Alicia M. Chandler[5], Eliza Wieman[4], Thomas R. Hopper[6], Nicolas E. Watkins[7], Jin Yue[1], Xinxin Cheng[8], Ming-Fu Lin[8], Duan Luo[8], Patrick L. Kramer[8], Xiaozhe Shen[8], Alexander H. Reid[8], Olaf Borkiewicz[1], Uta Ruett[1], Xiaoyi Zhang[1], Aaron M. Lindenberg[6,9], Jihong Ma[3,*], Richard Schaller[7,10*], Dmitri V. Talapin[3,10*], Benjamin L. Cotts[4*]

[1] X-ray Science Division, Argonne National Laboratory, Lemont, IL, 60527, USA.

[2] Department of Chemistry, James Franck Institute, and Pritzker School of Molecular Engineering, University of Chicago, Chicago, IL 60637, USA.

[3] Department of Mechanical Engineering, University of Vermont, Burlington, VT, 05405, USA.

[4] Department of Chemistry and Biochemistry, Middlebury College, Middlebury, VT, 05753, USA.

[5] School of Engineering, Brown University, Providence, RI, 02912, USA.

[6] Materials Science and Engineering Department, Stanford University, Stanford, CA, 94305, USA.

[7] Department of Chemistry, Northwestern University, Evanston, IL, 60208, USA.

[8] SLAC National Accelerator Laboratory, Menlo Park, CA 94025, USA.

[9] Department of Photon Science, Stanford University and SLAC National Accelerator Laboratory, Menlo Park, CA 94025, USA.

[10] Center for Nanomaterials, Argonne National Laboratory, Lemon, IL, 60527, USA.

*Correspondence to:

burakg@anl.gov, bcotts@middlebury.edu, schaller@anl.gov, dvtalapin@uchicago.edu, jihong.ma@uvm.edu

**Symmetry control is essential for realizing unconventional properties, such as ferroelectricity, nonlinear optical responses, and complex topological order, thus it holds promise for the design of emerging quantum and photonic systems. Nevertheless, fast and reversible control of symmetry in materials remains a challenge, especially for nanoscale systems. Here, we unveil reversible symmetry changes in colloidal lead chalcogenide quantum dots on picosecond timescales. Using a combination of ultrafast electron diffraction and total X-ray scattering, in conjunction with atomic-scale structural modeling and first-principles calculations, we reveal that symmetry-broken lead sulfide quantum dots restore to a centrosymmetric phase upon photoexcitation. The symmetry restoration is driven by photoexcited electronic carriers, which suppress lead off-centering for about 100 ps. Furthermore, the change in symmetry is closely correlated with the electronic properties as shown by transient optical measurements. Overall, this study elucidates reversible symmetry changes in colloidal quantum dots, and more broadly defines a new methodology to optically control symmetry in nanoscale systems on ultrafast timescales.**

Symmetry is a fundamental concept in physics, control of which bestows materials with unusual attributes, such as nonlinear optical responses[1], chirality[2], ferroelectricity[3], superconductivity[4], magnetoelectricity[5] and exotic quantum states[6]. Therefore, transient symmetry control, *i.e.,* changing symmetry on demand via external stimuli, is desirable for switchable functionalities in data storage, computing, and sensing[7,8]. Various materials allow symmetry control, including complex oxide perovskites[9–12], transition metal oxides[13] and metal chalcogenides[14–16], under various stimuli, including light[17], electric field[12], and pressure[18]. Among these, group IV – chalcogenides stand out, with structural instabilities resulting in symmetry breaking in their cubic rocksalt phases[19–21]. This behavior leads to unusually high dielectric polarizabilities in addition to large lattice anharmonicity, making IV-chalcogenides appealing for phase change, thermoelectric, and photonic applications[22].

Lead chalcogenides, a subgroup of group IV-chalcogenides, exhibit spontaneous symmetry breaking in their bulk form as shown by neutron and X-ray scattering[23–26]. This behavior is linked with a combination of effects, such as the stereochemical activity of lead lone pairs, resonant (or metavalent) bonding, and highly anharmonic interatomic potentials[19,24,25]. In their quantum-confined form, however, lead sulfide (PbS) quantum dots (QDs) exhibit permanent symmetry breaking via lead off-centering.[27,28] This is suggested to be induced by surface effects, such as strain heterogeneity imparted by surface ligands[27] and surface defects.[29]

From a technological point of view, PbS QDs are essential semiconductors used in a myriad of applications, including photodetectors[30], solar cells[31], light-emitters[32] and thermoelectrics[33]. Although optical properties can be controlled by surface chemistry, size, and shape[34–37], symmetry control under external stimuli has not been realized before. Particularly, photo-induced symmetry changes have remained completely uncharted in these QDs. Switching structural symmetry with light pulses is expected to empower new ways to engineer these QDs. For example, charge, exciton, and thermal transport can be dynamically manipulated, and chiral, nonlinear optical properties can be switched on and off in the QD devices.

Here, we study non-equilibrium symmetry changes in photoexcited PbS QDs via ultrafast diffraction techniques. We find that PbS QDs with a symmetry-broken ground-state structure transform into a centrosymmetric excited-state after ultrafast photoexcitation (Figure 1a). This response is reflected by the suppression of diffraction peaks associated with a lower symmetry lead off-centered phase, corroborated by atomistic modeling. Transient pair distribution function analysis shows that the Pb-S-Pb coordination changes via shift of Pb atoms, triggered by photogenerated carriers. The QDs revert to the symmetry-broken ground state as the photoexcited carriers are trapped or recombine. Finally, we study the effects of QD size, surface chemistry, and temperature on controlling the transient photoinduced symmetry changes.

**Ultrafast symmetry restoration in photoexcited PbS quantum dots**

To study transient symmetry changes in photoexcited PbS QDs, we perform ultrafast diffraction measurements using either femtosecond electron (Figure 1b) or picosecond x-ray pulses (Figure 1c) (Methods). First, we present electron diffraction measurements on PbS QDs capped with native oleic acid ligands (Figure S1). The results are not limited to a particular sample, as will be shown by measurements on QDs with varying size and surface chemistry. Optical pulses with an energy of 3.1 eV, 100 fs duration, and 360 Hz repetition rate, generate photocarriers in the QDs. Excitation fluence ranges from 0.35 to 1.0 mJ cm$^{-2}$, creating 15 to 50 electron-hole pairs per QD (photocarrier density of ~$10^{20}$ cm$^{-3}$) (Supplementary Note 1). Figure 2a shows the background-subtracted diffraction pattern, $I(Q)$, where $Q$ is the scattering vector (see Figure S2 before background subtraction). The Bragg peaks of the rocksalt PbS are also labeled in Figure 2a.

Structural changes induced by pulsed light excitation are encoded in a normalized difference pattern, $\Delta I(Q, t_{delay})/I_0(Q)$, where $t_{delay}$ is the time delay between the optical pump and electron probe pulses, and $I_0$ is the diffraction of the ground state. Figure 2b shows $\Delta I/I_0$ for delays of -2, 10, and 200 ps, while Figure 2c maps $\Delta I/I_0$ for all measured delays. A negative delay indicates the arrival of the electron pulse before photoexcitation, hence no transient effect. At 200 ps, the intensity of all Bragg peaks decreases, while the intensity between the Bragg peaks (i.e., diffuse scattering) increases (Figure 2b). This behavior implies heating of the QDs. Figures 2d-2e plot the Debye-Waller factor in $-\log\left(\frac{I}{I_0}\right)$ vs. $Q^2$. At 200 ps, $-\log\left(\frac{I}{I_0}\right)$ shows a linear response as a function of $Q^2$ (Figure 2e), confirming the transient heating[38,39]. Here, the induced mean-squared atomic displacement $\langle \Delta u^2 \rangle$ is estimated to be $5.5 \times 10^{-3}\text{Å}^2$, which translates into a temperature jump of 50 K (Supplementary Note 2). This is in agreement with a temperature jump of 37 K estimated from hot-carrier cooling effect (Supplementary Note 3).

At early time delays (i.e., 10 ps), additional features, *e.g.,* a dip at 2.6 Å$^{-1}$, emerge in $\Delta I/I_0$ (Figures 2b, 2c) and $-\log\left(\frac{I}{I_0}\right)$ deviates from the pure thermal (Debye-Waller) response (Figure 2d). Particularly, the dip at 2.6 Å$^{-1}$ does not match any Bragg reflection of the rocksalt PbS (Figure 2a). This $Q$ coincides with the nominally forbidden (211) peak, becoming allowed due to symmetry-breaking. We hypothesize that the symmetry-broken PbS QDs convert to a centrosymmetric excited state as the intensity of the lower symmetry (211) peak transiently decreases upon photoexcitation. While diffraction features associated with symmetry-breaking are hard to discern in $I_0(Q)$ due to broad QD diffraction peaks and subtle Pb off-centering (Figure S3), $\Delta I/I_0$ is sensitive to even small changes in the structural symmetry.

Dynamics measured at two different peaks (Figures 2f-2g) indicate two distinct processes, namely photoinduced heating and transient symmetry restoration. The response at 3.05 Å$^{-1}$, the symmetric (220) peak, rises with a lifetime of 1 ps and decays slowly beyond 0.5 ns (Figure 2f). These dynamics reflect photoinduced heating and subsequent thermal relaxation, respectively. Slow thermal relaxation can be explained by small thermal conductivity of these QDs[40], corroborated by our time-resolved XRD measurements (Figure S4 and Supplementary Note 4). The (220) peak seems to be immune to the particular symmetry changes with a flat response throughout the delay range examined (Figures 2c-2f). However, the response at 2.6 Å$^{-1}$, the lower-symmetry (211) peak, is receptive to the symmetry changes. Here, the signal emerges fast and decays quickly with lifetimes of 0.2 and 30 ps, respectively (Figure 2g). The initial rise is due to symmetry restoration, followed by symmetry recovery back to the ground state. When probing at 4.25 Å$^{-1}$, where symmetric (400) and lower-symmetry (322) peaks overlap, we simultaneously observe both slow (thermal) and fast (symmetry) recovery lifetimes (Figure S5).

The recovery timescales (Figures 2f, 2g) indicate that the symmetry restoration coexists with the thermal response up to 100 ps. Since the thermal relaxation is very slow, we calculate a difference diffraction pattern, $\Delta I_{sym}/I_0$, encoding only the symmetry restoration. For this, we subtract the later time heating response, $I(Q, 500\ ps)/I_0$, from that of the earlier mixed response, $I(Q, 5\ ps)/I_0$. To model $\Delta I_{sym}/I_0$, we simulate QD diffraction patterns using the Debye scattering equation (Methods). The simulated QD diffraction pattern (Figure 2h) agrees well with the experimental $I_0(Q)$, except slight peak height deviations due to preferred-orientation in the QD films. We calculate $\Delta I_{model}/I_0$ by modeling various structural and symmetry changes, including changes in strain, disorder, defect creation/annihilation, as well as lead off-centering restoration (Methods). $\Delta I_{model}/I_0$ shows close agreement with $\Delta I_{sym}/I_0$ (Figure 2k) only when the modeled QDs transition from lead off-centering to no-off-centering. Off-centering restoration decreases the intensity of particular lower-symmetry peaks, agreeing with the experimental data. Simulations with other perturbations deviate from $\Delta I_{sym}/I_0$ (see Figures S6-S8). Therefore, the atomistic structural model provides strong support for the transient symmetry restoration hypothesis.

To gain deeper insights into the transient symmetry recovery, we consider lead off-centering restoration along different crystallographic directions (e.g., <100>, <110> and <111>). We obtain the smallest deviation between the experiment and model ($\Delta_{residual}$) for the restoration of lead off-centering along <110> with a total displacement amplitude of 0.07 Å (Figure S9). The amplitude of lead off-centering recovery is in line with the static lead off-centering that is 0.05 – 0.07 Å for PbS QDs with a diameter ranging from 3 to 7 nm[28]. To note, the transient displacement of lead atoms due to off-centering restoration is also comparable to the photoinduced thermal displacements, $\sqrt{\langle \Delta u^2 \rangle} \approx 0.075$ Å, induced by 1 mJ cm$^{-2}$ excitation.

Next, we study the temperature dependence of the transient off-centering restoration (Figure 2l). The transient symmetry response disappears below 200 K and reversibly comes back at room temperature (Figure S10). This behavior indicates that symmetry changes, reflected by a signature dip at 2.6 Å$^{-1}$, cannot be due to impurities, such as lead nanoparticles having a Bragg peak at this position. Otherwise, the impurity signal would persist at all temperatures. Thus, photoexcited transient symmetry restoration must be an intrinsic response of the PbS QDs. Supporting this, data from an earlier ultrafast diffraction measurement showed the dip at 2.6 Å$^{-1}$ in PbS QDs synthesized at a different lab[41], although the origin of this feature was never discussed.

The disappearance of photoinduced symmetry restoration at low temperatures may be due to ligand shell solidication[42], which may alter the QDs' strain state. Thus, PbS QDs may become more centrosymmetric at low temperatures (Figure S11). The increase in symmetry at low temperatures is also akin to the bulk lead chalcogenides, where local symmetry breaking has been observed only at elevated temperatures. This behavior was attributed to order-disorder transition or "emphanisis" effect, causing spontaneous symmetry-breaking upon warming of the lattice.[23,24]

We measure transient symmetry changes down to an excitation fluence of 0.35 mJ cm$^{-2}$ (Figure S12a), creating 15 electron-hole pairs per QD. Due to experimental limitations, we could not test lower fluences. However, fluence-dependent trend shows a saturation behavior (Figure S12b). Thus, symmetry restoration may occur even under weaker excitation densities. Moreover, symmetry restoration is confirmed when exciting with a lower photon energy (see 2 eV excitation in Figure S13). Thus, transient symmetry changes persistently occur in the photoexcited PbS QDs.

**Driving force behind transient symmetry restoration**

To understand the driving force behind reversible, photoinduced symmetry changes, we perform transient absorption measurements (Methods). These measurements probe the transient electronic processes that accompany the symmetry changes. The transient absorption data shows a ground state bleach of the QD band gap due to the photogeneration of electron-hole pairs (Figure 3a). The bleach signal recovers with multiple time constants, the fastest one being ~30 ps (Figure 3b). This fast recovery lifetime does not exhibit fluence dependence for the pump intensity range examined (Figure S14). Therefore, the initial carrier population decay is due to trapping rather than exciton-exciton annihilation[43]. The lifetime of the initial population decay matches well with that of symmetry recovery (30 ps) (Figure 2g). This agreement indicates that the photogenerated electron-hole pairs trigger the transient symmetry changes in the PbS QDs.

First-principles calculations show a hint of symmetry improvement under electron-hole excitation (Supplementary Note 5). We model a small PbS cluster of 64 Pb-S atoms, where the variance of Pb – S bond angle ($\sigma^2$) in the central part of the cluster reduces by 11%, from $8.9 \times 10^{-5}$ (in $rad^2$) to $7.9 \times 10^{-5}$, upon introducing 0.5 electron-hole pairs (density of ~$5 \times 10^{19}$ cm$^{-3}$). The decrease in $\sigma^2$ around a nominal bond angle of 90° points out to an improvement of the structural symmetry toward the cubic phase after introducing electron-hole pairs.

To gain insights into the impact of symmetry changes on the electronic properties, we compare the transient absorption at 100 K vs. 300 K. At 100 K, the bleach red-shifts (9 meV) within a few ps after photoexcitation. This can be understood by carrier-carrier screening effect[44]. At 300 K, in addition to a similar red-shift, we also observe a low energy shoulder building up around 2000 nm, which is separated from the main bleach by ~40 meV. The shoulder cannot be explained by screening, but it can be via formation of a QD sub-population with a smaller bandgap. We hypothesize that the symmetry-restored QDs are responsible for this behavior. First-principles calculations of PbS bandgap with varying lead off-centering displacements (Methods) show that decreasing off-centering reduces the bandgap (inset of Figure 3d, Figure S15). For an off-centering restoration of 0.07 Å, the estimated change in bandgap is -20 meV. Thus, smaller bandgap of the symmetry-restored QDs may concur with the low-energy shoulder formation. Furthermore, Figure 3d overlays the recovery dynamics of the low energy shoulder vs. symmetry restoration. The close temporal agreement highlights the strong coupling between the structural symmetry and the electronic properties.

**Atomistic mechanism of symmetry restoration**

To investigate how lead off-centering transiently diminishes under photoexcitation, we perform transient atomic pair distribution function analysis (Methods). The pair distribution function, $G(r)$, represents the probability of finding a pair of atoms separated by a distance $r$. We calculate both static $G(r)$ (Figure 4a) and transient $\Delta G(r, t_{delay})$ at varying pump-probe delays (Figure 4b-c), and focus on the changes in the short-range order (<5 Å). $G(r)$ shows the first peak at 2.9 Å due to the smallest Pb-S bond length. The second peak at 4.1 Å is due to the nearest Pb-Pb (and S-S) distance. When the QDs are symmetry-restored at 10 ps (Figure 4b), $\Delta G$ shows no detectable change for the first pair, while the second pair is broadened. This response is corroborated by time-resolved total x-ray scattering measurements (Figure 4c). Broadening of the Pb-Pb pair without changing the Pb-S bond length suggests a change in Pb-S-Pb coordination, arising from correlated shift of Pb atoms. Using x-rays, we also compare photoexcitation to heating (Figure 4c). Under static heating, $\Delta G$ shows broadening of both Pb-S and Pb-Pb pairs due to isotropic thermal disorder, in accordance with molecular dynamic simulations (Figure S16 and Supplementary Note 6). To note, the QDs are provided in a liquid jet for x-ray vs. solid film for electron diffraction. Thus, the symmetry changes seem to occur independent of the sample form.

To correlate photoinduced symmetry changes with atom position changes, we consider net atom displacements that can be induced by various phonons. For this, we calculate the phonon dispersion of PbS (Methods) with lead off-centering (Figure 4d and Figure S17), and consider low energy modes at 1.3, 1.6 and 2 THz. The mode at 1.3 THz is a dispersionless transverse acoustic (TA) phonon[45], which involves localized lead atom motion within the lead sublattice (Figure 4e). 1.6 THz mode is a longitudinal acoustic (LA) phonon, involving coupled lead and sulfur motion. 2 THz mode is a transverse optical (TO) phonon. We calculate $\Delta G(r)$ induced by a net displacement along each phonon's coordinate (Methods). Net displacement (amplitude 0.05 Å) along the TA mode broadens the second pair (Pb-Pb) without changing the first (Pb-S) (Figure 4e). On the other hand, atom displacements along LA and TO modes broaden the first pair (Pb-S). Thus, the shift of Pb atoms via the TA mode can explain the transient off-centering restoration. The rise time of symmetry restoration also agrees with the TA mode. A net displacement arises during phonon's quarter period, which locks the structure to its new symmetry state[46]. The rise time (0.2 ps) of symmetry restoration concurs with the quarter period (0.19 ps) of the TA mode.

**Controlling transient symmetry restoration via QD size and surface chemistry**

To gain control on the transient symmetry responses in PbS QDs, we investigate the effect of QD size and surface chemistry. To compare symmetry changes among different samples, we define a figure of merit, which is the absolute, normalized difference diffraction at the lower-symmetry (211) peak. $\left|\Delta I/I_0\right|_{(211)}$ encodes the transient Pb off-centering restoration.

First, we investigate QDs with diameters of 5.7, 7.3 and 8.1 nm, having either organic oleic acid (OA) or inorganic $Na_4Sn_2S_6$ ligands (Methods). While the QDs with OA ligands show much stronger transient symmetry changes, the transient symmetry restoration becomes stronger for the smaller QD size for both ligands (Figure 5a). This size effect is consistent with Ref.[27] that the static symmetry-breaking is more pronounced in smaller QDs. To study ligand effect, we keep the QD size the same (7.3 nm) and vary the surface capping. We test -SH, -OA, -CdS/OA, -$Cl_2$ and -$Sn_2S_6^{4-}$ ligands and surface treatments (Methods). The transient symmetry change (Figure 5b) is the strongest with the -OA, which preferentially bind to <111> facets[47] and cause strain heterogeneity in the QDs[27]. The QDs with -$Cl_2$ and -SH treated surfaces show measurable but smaller photoinduced symmetry changes. On the other hand, the QDs with inorganic -$Sn_2S_6^{4-}$ ligand and thin CdS shell show suppressed symmetry changes. $\Delta G(r,t)$ of the PbS QD capped with -$Sn_2S_6^{4-}$ also confirm the thermal-like photoexcitation response (Figure S18). The suppression in symmetry changes may be due to weaker static symmetry-breaking with particular surface chemistries. Overall, surface chemistry can be used to manipulate photoinduced symmetry changes.

**Discussion and conclusion**

Ultrafast structural experiments combining atomistic structural modeling elucidate that the symmetry-broken PbS QDs undergo reversible, picosecond timescale symmetry changes upon photoexcitation (Figure S19). These symmetry changes, underpinned by the restoration of lead off-centering displacements, strongly couple with the electronic, hence optical, properties. Therefore, photoinduced, ultrafast symmetry changes can be exploited to manipulate excited state functionalities of the QDs, such as charge, exciton and thermal transport.

We show that the symmetry-restored QDs have a red-shifted bandgap (Figure 3c), which could be used to control exciton transport by harnessing the symmetric QDs as exciton traps. Recently, Ref. [48] showed that exciton diffusion in photoexcited PbS QD films is fast initially (< 0.5 ps), which then slows down to become sub-diffusive. The transition time in exciton diffusivity coincides with the onset of the symmetry restoration. Thus, the symmetry restoration of the QDs may explain dramatic transient changes in exciton diffusivity in the QD solid thin films.

Dynamic, on-demand manipulation of QD properties may open up new avenues for stimuli-controlled photonic, solar and quantum devices, including control of chiral properties[49], catalytic activity and lasing thresholds[50]. We anticipate that transient symmetry changes may be realized under weaker, yet more affordable optical excitation, as well as electrical and electrochemical stimulations. To this end, our combined experimental and modeling approach defines a comprehensive methodology to study transient symmetry changes in QDs, which can be applied to broader nanocrystalline systems. Future work will focus on engineering timescales of symmetry changes, harnessing different excitation mechanisms and developing adaptive controls in QD devices.

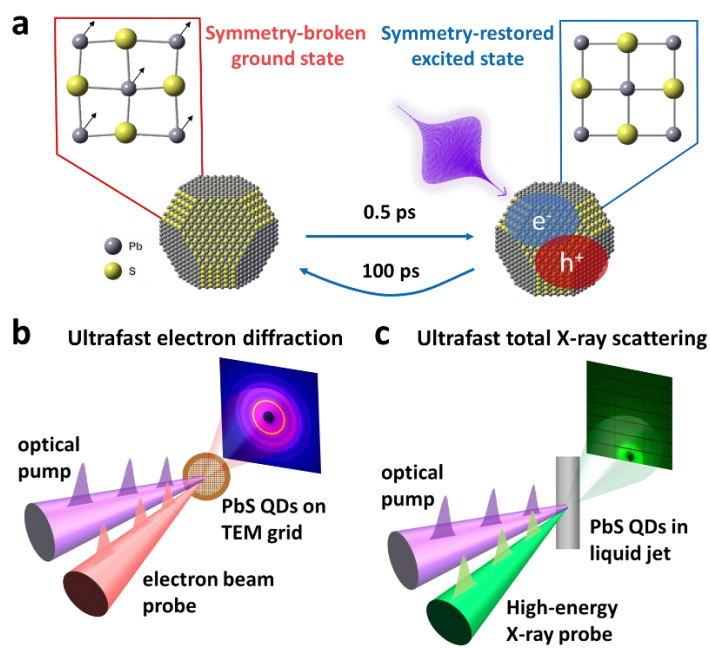

**Fig. 1 | Symmetry control in photoexcited lead sulfide quantum dots (QDs). (a)** Schematic showing symmetry-broken PbS quantum dots (QDs) in the ground state vs. symmetry-restored QDs in the excited state. Schematics of the experimental techniques used to study transient structural symmetry changes via (**b**) ultrafast electron diffraction, and (**c**) ultrafast total X-ray scattering.

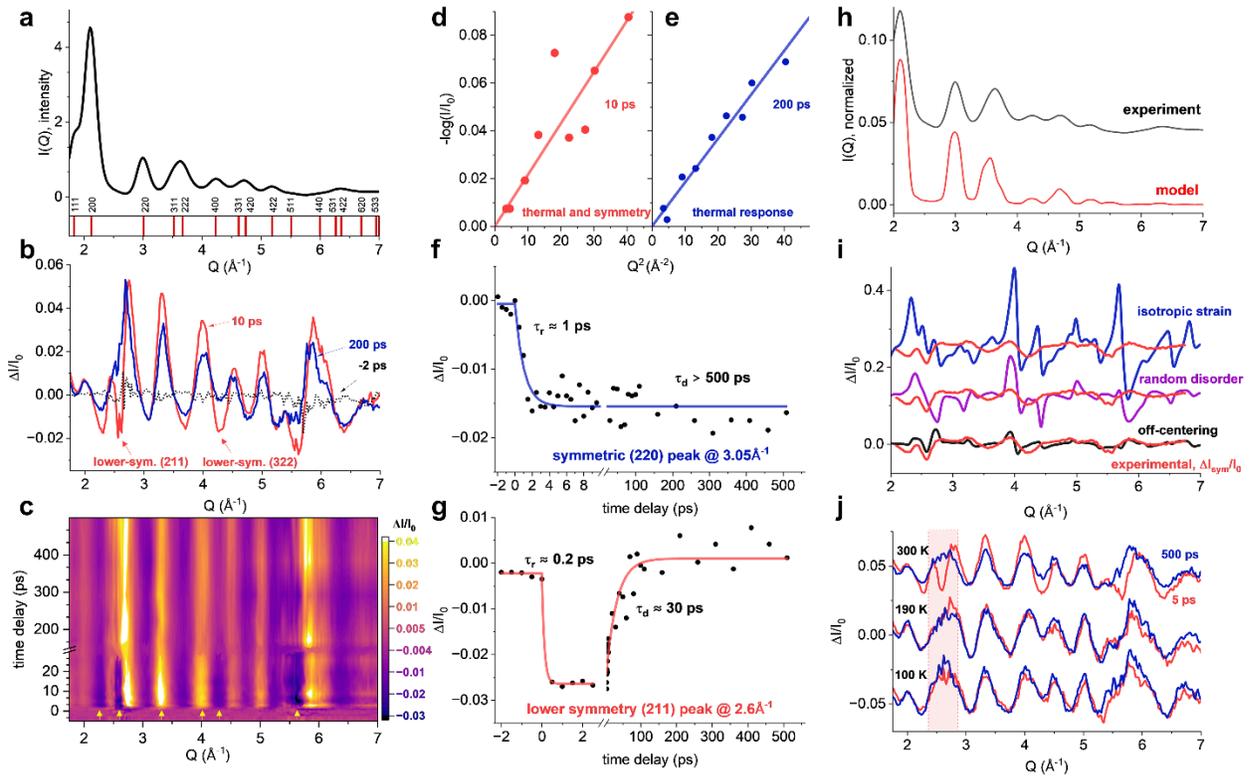

**Fig. 2 | Structural response in photoexcited PbS QDs.** (**a**) Experimental static diffraction pattern, $I(Q)$, where $Q$ is scattering vector. $I(Q)$ is subtracted for background signal. Bragg peaks of the rocksalt PbS phase are labeled in the bottom panel. (**b**) Difference diffraction pattern normalized by static diffraction, $\Delta I/I_0$, presented at -2 ps (black dots), 10 ps (red) and 200 ps (blue). The position of symmetry forbidden (211) and (322) peaks are labeled. (**c**) $\Delta I/I_0$ as a function of $Q$ vs. time delay. The color bar encodes the normalized difference signal changes. Yellow arrows at the bottom indicate the regions with fast initial recovery associated with symmetry change. (**d-e**) Debye Waller factor presented as $-\log\left(\frac{I(Q)}{I_0(Q)}\right)$ vs. $Q^2$ measured at pump-probe delays of 10 ps (**d**) and 200 ps (**e**). Rise and decay dynamics at the symmetric (220) peak (**f**), and the lower symmetry (211) peak (**g**). The rise ($\tau_r$) and decay ($\tau_d$) lifetimes are noted in each panel. (**h**) Experimental diffraction pattern vs. calculated diffraction pattern by the Debye scattering equation model. (**i**) Experimental $\frac{\Delta I_{sym}}{I_0}$, encoding only symmetry changes, vs. modeled $\frac{\Delta I}{I_0}$ for isotropic strain, random disorder and off-centering displacements. (**j**) $\Delta I/I_0$ at 5 ps at temperatures of 300 K, 190 K and 100 K. The region around 2.6 Å$^{-1}$ is highlighted by a transparent red rectangle.

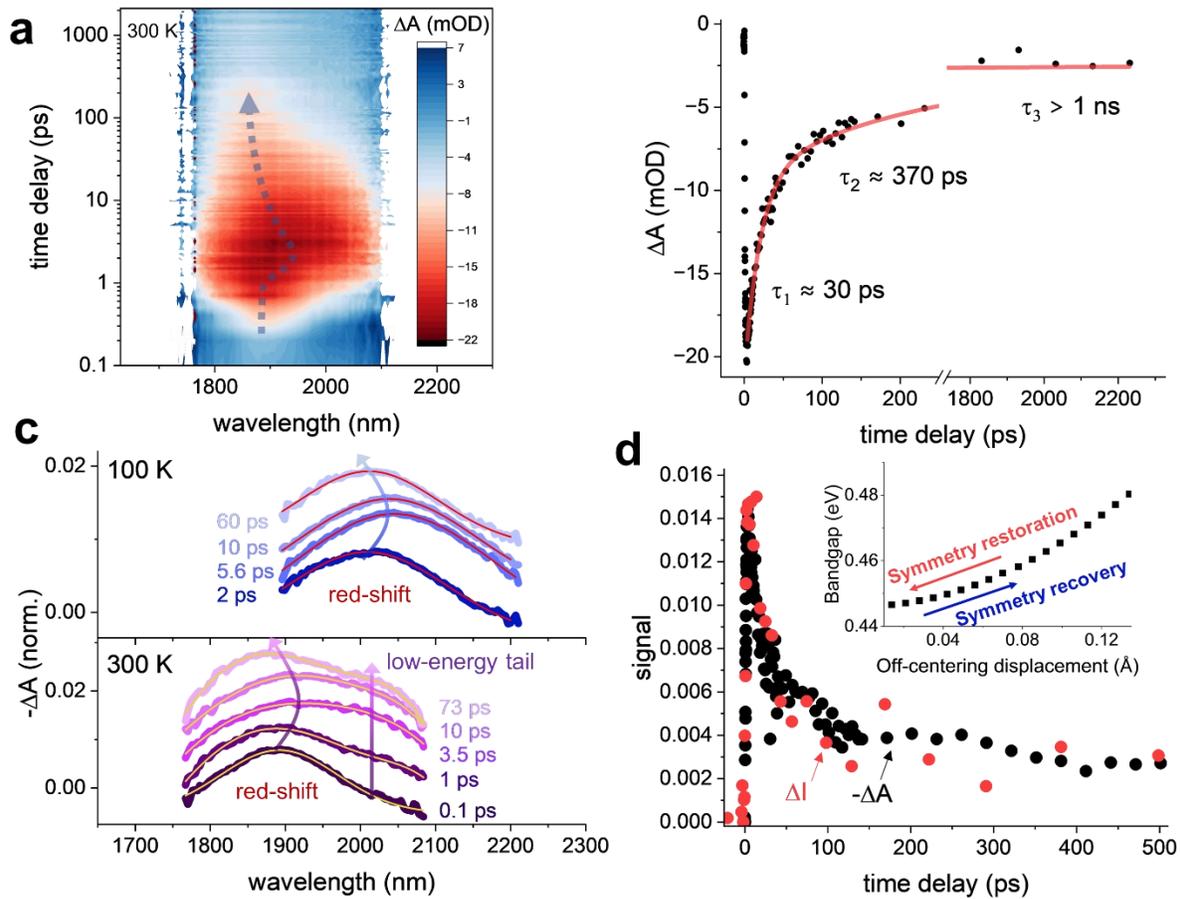

**Fig. 3 | Measurement of transient electronic responses.** (**a**) Transient absorption measurements in 8.3 nm PbS QDs (oleic acid ligands) around its band edge at 1850 nm at 300 K. Color bar encodes the change in absorption ($\Delta A$ in milli optical density, mOD). Negative changes indicate ground state bleach by photogenerated electron-hole pairs. (**b**) Transient absorption data at the band edge shows the transient recovery of the absorption with three decay lifetime components: 30 ps, 370 ps and > 1 ns. (**c**) Transient bleach profiles measured at 100 K (top panel) and 300 K (bottom panel). 100 K data shows a slight red-shift of the profile which recovers within 60 ps. The red-shift arises from carrier-carrier screening. 300 K data shows a similar red-shift together with a significant broadening at the low-energy tail around 2000 nm. (**d**) Low-energy tail (change in absorption, $-\Delta A$) vs. symmetry recovery (transient diffraction signal $\Delta I$ at 4.25 Å$^{-1}$) as a function of delay. Close temporal agreement shows strong correlation between the excited-state bandgap vs. structural symmetry. The inset of (**d**) shows first-principles calculation of the PbS bandgap vs. lead off-centering displacement amplitude. Symmetry restoration reduces the bandgap.

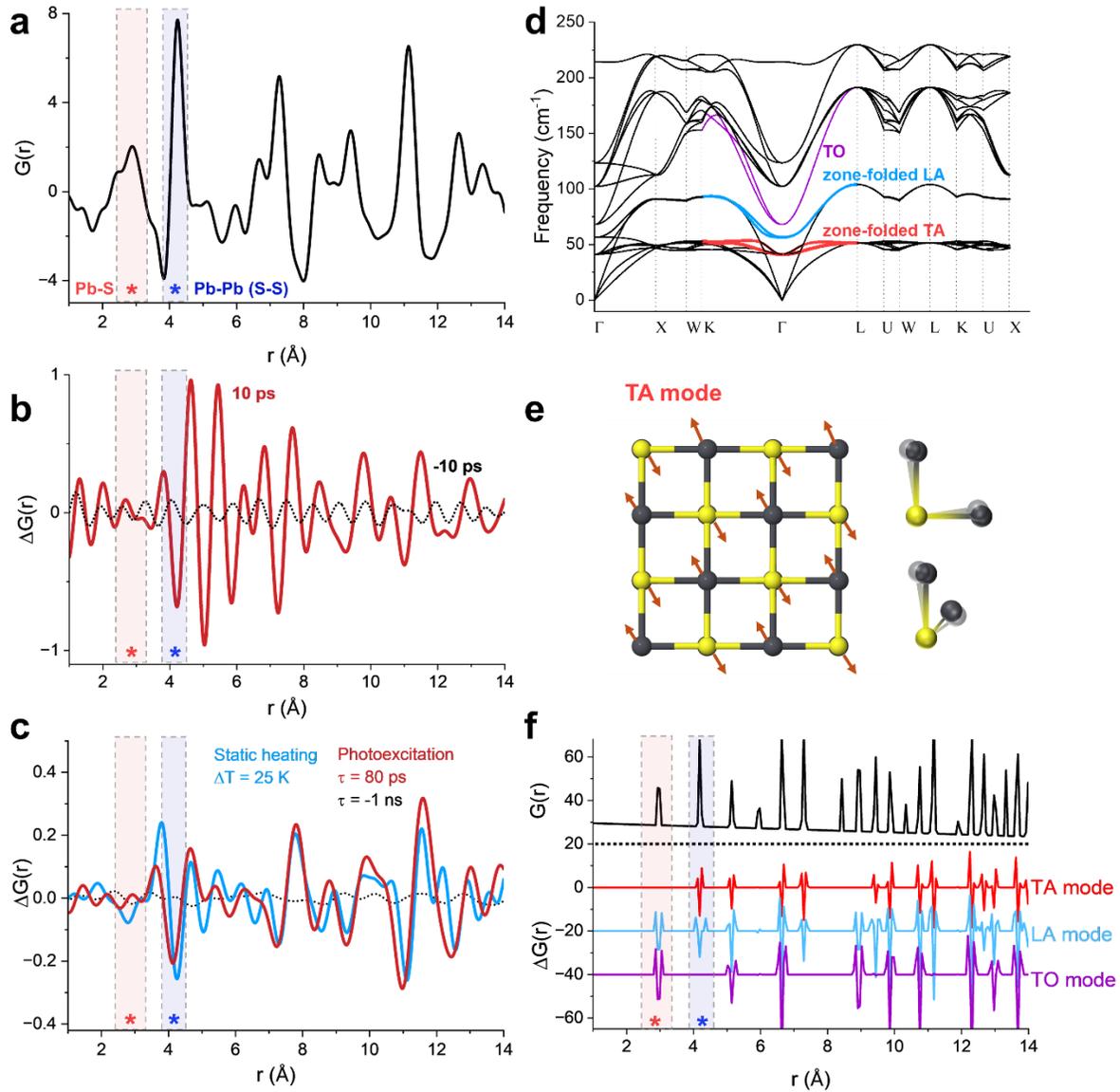

**Fig. 4 | Atomistic mechanism of symmetry restoration.** (**a**) Atomic pair distribution function, $G(r)$, measured for PbS QDs. Peaks at 2.9 and 4.1 Å are due to the shortest Pb-S and Pb-Pb (S-S) pair distances. Stars symbols at the bottom marks the positions of Pb-S (red) and Pb-Pb (blue) pairs. (**b**) Transient pair distribution function, $\Delta G(r)$, at 10 ps measured by ultrafast electron diffraction. Noise level is shown by the transient data measured at -10 ps. Pb-S pair (2.9 Å) does not show a measurable change while Pb-Pb pair (4.1 Å) gets broadened. (**c**) $\Delta G(r)$ measured by time-resolved total X-ray scattering at a delay of 80 ps (red) and -1 ns (black, noise level), in addition to static heating (blue) of $\Delta T = 25°C$ above room temperature. (**d**) Phonon dispersion of lead off-centered PbS structure. Low energy transverse acoustic (TA), longitudinal acoustic (LA) and transverse optical (TO) modes are marked with red, blue and purple

colors, respectively. (**e**) The TA phonon mode involves correlated lead shifts within the off-centered lead sublattice without sulfur motion. (**f**) Change in pair distribution function, $\Delta G(r)$, induced by a net displacement along either TA, LA and TO phonon coordinates with a maximum displacement amplitude of 0.05 Å. $\Delta G(r)$ induced by the TA mode shows close resemblance to the photoinduced $\Delta G(r)$ response.

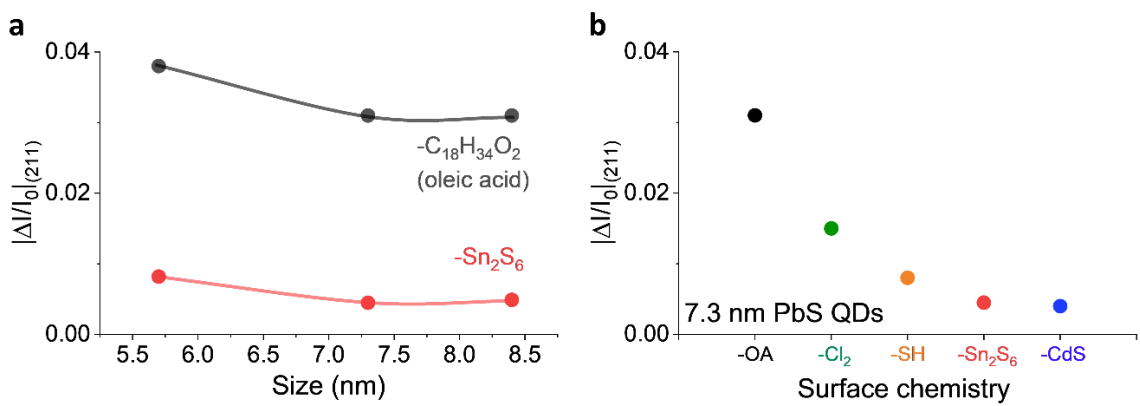

**Fig. 5 | Effects of quantum dot size and surface chemistry on transient symmetry restoration.** Normalized difference diffraction intensity at the lower-symmetry (211) peak for different QD sizes (**a**) and varying surface chemistry with the same QD size (7.3 nm) (**b**).

**Methods**

**Synthesis of PbS quantum dots.**

PbS QDs with OA ligands (PbS-OA) were synthesized in accordance with a previously reported procedure.[51] The PbS QDs were washed four times using a toluene/methyl acetate mixture before being redispersed in toluene and stored under a $N_2$ atmosphere.

PbS QDs with inorganic $Na_4Sn_2S_6$ ligands (PbS-$Sn_2S_6$) were prepared through the substitution of OA ligands with $Na_4Sn_2S_6$ ligands. Initially, a 0.20 M solution of $Na_4Sn_2S_6$ was prepared by stirring $Na_2S$ (312 mg, 4 mmol) and $SnS_2$ (731 mg, 4 mmol) in N-methylformamide (NMF) (10 ml) overnight. Residual solids were removed by centrifugation. Then, a biphasic mixture consisting of 25 mM $Na_4Sn_2S_6$ salts in NMF and PbS QDs with OA ligands in n-hexane was prepared and stirred overnight. Following stirring, the PbS QDs were transferred from the n-hexane phase to the NMF phase, resulting in PbS QDs with $Na_4Sn_2S_6$ ligands. The n-hexane layer was subsequently replaced with fresh hexane three times to eliminate dissociated oleate ligands. The PbS QDs in NMF were then precipitated by the addition of acetonitrile and were redispersed in NMF for storage under a $N_2$ atmosphere.

Chloride-treated QDs (PbS-PbCl$_2$/OAm) were prepared by the following procedure. Initially, PbS-OA QDs were dispersed in a solution of 10 mM OAm in hexane. Then, ~10 mM of PbCl$_2$ was added and suspended in this solution. The solution mixture was mixed vigorously for 5 minutes to facilitate the ligand exchange. Following the ligand exchange, the excess PbCl$_2$ was removed by centrifugation and removal of the solid residual. The excess OAm and OAc were removed by precipitating the PbS QDs with ethanol and then redispersed in toluene for storage under a N$_2$ atmosphere.

PbS QDs with PbS-CdS-OA core-shell structure (PbS-CdS) were prepared by adding a CdS shell according to a previously reported procedure.[52]

Thiol-treated PbS QDs (PbS-SH) were prepared by vigorously mixing a solution of PbS-OA QDs with ~10 % v/v dodecanethiol for 5 minutes. The treated PbS QDs were precipitated with ethanol and then redispersed in toluene for storage under a N$_2$ atmosphere.

To prepare samples for the MeV-UED experiments, we drop casted diluted samples onto a TEM grid with ultrathin amorphous carbon support. For thin film samples, we spin coated to produce thin films. For time-resolved total x-ray scattering, the QDs with OA ligands were dissolved in dodecane in high concentration. In high concentration solution, the optical density of the mixture at 3.1 eV was ~6.

**Ultrafast electron diffraction (UED).**

UED measurements were performed at the MeV-UED facility at the SLAC National Laboratory[53]. To generate optical pump and electron probe beams, output of a multipass Ti:sapphire laser (800nm, 60 fs, 360 Hz) is split into two. Frequency tripled (267 nm) beam excites the photocathode to produce electrons, which are then accelerated to 3.5 MeV (~150 fs FWHM) for 20 fC per pulse using a Klystron. The size of the electron beam on the sample was ca. 100 μm. For the optical excitation beam, we frequency doubled the fundamental output to achieve 400 nm with a pulse width of 75 fs, a repetition rate of 360 Hz. The optical pump beam size was 450 μm on the sample location. The optical pump beam was quasi collinear with the electron probe beam. The diffraction from the samples were collected in transmission using a red phosphor screen and Andor iXon Ultra 888 EMCCD camera. Pump-probe time delays were controlled by a mechanical delay stage. We calibrated time zero using bismuth and confirmed the spatial overlap using YAG:Ce. To minimize systematic errors, we randomized delay points in every time delay scan.

**Ultrafast total X-ray scattering.**

These measurements were performed at the Beamline 11-ID-D of the Advanced Photon Source. We excited the samples with a frequency doubled output (400 nm) of a Ti:Sa laser with 10 kHz repetition rate and 1.7 ps pulse width. The size of the beam was 250 $\mu$m (V) × 1000 $\mu$m (H) on the sample location. As a probe, we used an X-ray energy of 17.5 keV, 80 ps pulse width and 6.5 MHz repetition rate. The size of the X-ray probe was 50 $\mu$m (V) × 450 $\mu$m (H) on the sample location. The PbS QD (OA ligands) were dissolved in dodecane in high concentration (optical density of 6 for 1 mm optical path). The QD sample was circulated as a round liquid jet with a diameter of 750 $\mu$m. We collected the total scattering with a Pilatus 2M detector with 300 $\mu$m thick silicon active area and pixel size of 172 by 172 $\mu$m. We used $CeO_2$ calibrant to calibrate the sample-detector distance (17.3 cm). The maximum measured $Q$ range was 8.5 Å$^{-1}$. We electronically gated the Pilatus detector at the repetition rate of the laser to collect pump-probe signal. We performed temporal overlap using a fast rise time photodiode. Spatial overlap is checked with a 300 $\mu$m pinhole at the sample location. We controlled the pump-probe delay electronically by a phase shifter (Colby Instruments) and a delay generator (Highland V85x). For each delay point, we acquired data for 5 sec. We randomized the delay points to minimize systematic errors. We repeated the time delay scans by 5 to 8 times to obtain improved signal to noise. We converted total scattering images to diffraction pattern vs. $Q$ by using QXRD software.

**Debye Scattering Equation Modeling.**

We use open-source command line programs (ATOMSK[54] and DebyeByPy[55]) through Jupyter Notebook to perform Debye scattering equation modeling of the PbS QDs. In ATOMSK, we create a PbS supercell 6 nm in edge length. Then, we import a cuboctahedron shape and use it to mimic surface faceting, defining a QD with a diameter of 6 nm. On the ATOMSK command line in Jupyter Notebook, we model symmetry changes, strain effects, random disorder, and dislocations for this QD using built-in ATOMSK functions. We loop through distortions with a specified range of displacement amplitudes, strain percentages, or Burgers vectors. For symmetry changes associated with lead off-centering, we select all lead atoms within the QD. Then, we use the ATOMSK "shift" function to shift positions of the selected lead atoms along different (<100>, <110> and <111>) crystallographic directions with amplitudes from 0.01 Å to 0.09 Å (along each axis within that crystallographic direction). For strain effects, we apply the "deform" function with 0.01 to 0.1% strain along various directions. For generating random deformations, we use the "disturb" function to randomly move the positions of atoms from 0.01 to 0.15 Å. For modeling dislocations, we loop through Burgers vectors from 0.1 to 0.9 Å (Figure S8). We automatically save each set of simulated QD structure as .xyz files in the same folder. Files in this folder are then automatically loaded into DebyeByPy to produce the respective $I(Q)$ diffraction patterns. DebyeByPy inherently utilizes the Debye scattering equation to calculate the diffraction pattern of the QDs and considers atomic form factors of Pb and S, as well as the Debye Waller factor. To mimic instrumental $Q$-broadening (0.23 Å$^{-1}$) in the UED experiments, we applied a square shaped convolution function to broaden the peaks. We use Jupyter Notebook to obtain $\Delta I(Q)/I_0$ by subtracting the simulated diffraction pattern of distorted QDs from the bare QD then dividing by the distorted QD. Here, distortion again means either lead off-centering, strain, random deformations, or dislocations. Then, we compared different models by calculating;

$$\Delta_{residual} = \Sigma_Q \left( \left(\frac{\Delta I(Q)}{I_0}\right)_{sym.} - \left(\frac{\Delta I(Q)}{I_0}\right)_{model} \right)^2.$$

**Transient absorption measurements.**

Transient absorption measurements were performed at the Center for Nanoscale Materials (CNM) at Argonne National Laboratory. PbS QDs (8.3 nm in diameter capped with OA ligands) were drop casted on substrates transparent in the infrared range. A Ti:Sa laser (2 kHz) output was divided into two. One arm was directed to an optical parametric amplifier to generate near-infrared probe photons near the temperature-dependent band edge of the QDs (centered near 1800 nm for 298 K measurements and near 2000 nm for 100 K measurements) using the idler beam. One arm frequency doubled the laser's fundamental wavelength to produce 400 nm pump pulses that were mechanical chopped to reduce the repetition rate to 1 kHz. A red-extended, InGaAs array detector was used to measure each probe pulse spectrum following dispersion in a spectrograph with shot rejection monitoring to increase signal to noise. Pump intensity was adjusted using a continuously variable neutral density filter. The TA data was corrected for a chirp in the probe pulse using the software Surface Xplorer. For fluence dependent TA measurements, we used the PbS QDs dissolved in toluene (Figure S14).

**First-principles calculations.**

First-principles calculations were performed using the Vienna Ab initio Simulation Package (VASP)[56]. The exchange-correlation potential was defined at the level of the generalized gradient approximation (GGA) using the Perdew–Burke Ernzerhof (PBE) functional[57,58]. The projected augmented wave (PAW)[59] method was employed to describe the electronic wave function with an energy cutoff of 500 eV. The calculations were carried out in a $2 \times 2 \times 2$ PbS supercell, and the Brillouin zone was integrated with a $4 \times 4 \times 4$ Γ-centered k-mesh for geometry optimization.

We examined the changes in the band gap with Pb-off-centering displacements ranging from 0.0017 to 0.2Å along the <111>, <110> and <100> directions.

To perform phonon dispersion calculations, we used the Phonopy package[60], considering the analytic correction on the PbS bulk structure using a 2 × 2 × 2 supercell after the system was structurally relaxed via VASP. Then, phonon dispersion curves for bulk pristine and displaced systems (Pb-off-centering displacement of 0.07 Å along the <110> direction) are calculated.

To calculate changes in pair distribution function $G(r)$ due to a net displacement along one of the phonon eigen directions, we further expanded the supercell with modal displacements by ten times along the $x$-, $y$-, and $z$-directions. The magnitude of the mode shapes is scaled by a displacement amplitude of 0.05 Å. We first calculate the radial distribution function (RDF) of the expanded supercell describing the likelihood of finding a neighboring atom in the spherical shell of a central atom, $g(r)$

$$g(r) = \frac{n(r)}{\rho_o 4\pi r^2 \Delta r},$$

Where $n(r)$ is the number of atoms in the spherical shell with a radius $r$ from the central atom with a thickness of $\Delta r$, $\rho_o$ is the number density. $G(r)$ then is computed through

$$G(r) = 4\pi\rho_o r[g(r) - 1].$$

**Time-resolved atomic pair distribution function analysis.**

We calculate change in atomic pair distribution function, $\Delta G(r,t)$ to examine the transient evolution of the atomic pair correlation function ($G(r)$) as a function of pump-probe delay ($t$). We calculate the difference diffraction intensity $\Delta I(Q,t)$ which is $I(Q,t) - I_0(Q)$. $I_0(Q)$ is the diffraction intensity before the laser arrives (i.e., ground-state). Difference total structure function is then estimated: $\Delta S(Q,t) = \frac{\Delta I(Q,t)}{|f|^2}$, where $f$ is the atomic form factor that is represented as:

$$f = \sum_j a_j e^{(-b_j s^2)} + \frac{m_0 e^2}{8\pi^2 \hbar^2}\left(\frac{\Delta Z}{s^2}\right)$$

To calculate $f$, we obtained a and b coefficients from an earlier report, we related $Q$ and $s$ as $s = \frac{\sin(\theta)}{\lambda} = \frac{Q}{4\pi}$. To calculate the change in (or difference) PDF, we take sine Fourier transform as $\Delta G(r,t) = \frac{2}{\pi}\int_{Q_{min}}^{Q_{max}} Q * \Delta S(Q,t) * \sin(Qr)\, dQ$.

**Static atomic pair distribution function measurements.**
We used Beamline 11-ID-B at the Advanced Photon Source to measure static $G(r)$ of PbS QDs. We used 86.5 keV x-ray energy. PbS QDs were drop casted on quartz films. Bare quartz substrate measurements were also done. We used the software GSASII to reduce data to a one-dimensional diffraction pattern and then another software PDFGetX2 to convert the total scattering data into $G(r)$. Sample was heated with a ceramic heater. Sample height with respect to x-ray beam was recalibrated after temperature changes.

**Acknowledgments:** Work performed at the Center for Nanoscale Materials and Advanced Photon Source, both U.S. Department of Energy Office of Science User Facilities, was supported by the U.S. DOE, Office of Basic Energy Sciences, under Contract No. DE-AC02-06CH11357. MeV-UED is operated as part of the Linac Coherent Light Source at the SLAC National Accelerator Laboratory, supported by the U.S. Department of Energy, Office of Science, Office of Basic Energy Sciences under Contract No. DE-AC02-76SF00515. S.G., M.T., E.W., J.M., and B.L.C. acknowledge support by the U.S. Department of Energy, Office of Science, Office of Basic Energy Sciences under Award Number DE-SC0023425. A. J. is partially supported by Kwanjeong Educational Foundation. D. V. T. also acknowledges support by the National Science Foundation under award number DMR-2019444. This work made use of the shared facilities at the University of Chicago Materials Research Science and Engineering Center, supported by National Science Foundation under award number DMR-2011854. N.E.W. and R.D.S. acknowledge support of the National Science Foundation under MSN-1808950. This work was supported in part by the U.S. Department of Energy, Office of Science, Office of Workforce Development for Teachers and Scientists (WDTS) under the Science Undergraduate Laboratory Internships Program (SULI).

**Author contributions:**

**Competing interests:** Authors declare no competing interests.

**Data Availability:** The datasets generated and/or analysed during the current study are available from the corresponding author on reasonable request.